\documentclass[10pt]{iopart}
\usepackage[dvipdfmx]{graphicx}
\usepackage{ulem}
\usepackage{here}
\def\vector#1{\mbox{\boldmath $#1$}}

\begin{document}

\title{A ``mean-field approximation'' on the phase transitions of three-dimensional Lennard-Jones model }
\author{Hisato Komatsu}
\address{ Graduate school of Arts and Sciences, The University of Tokyo, 3-8-1 Komaba, Meguro-ku, Tokyo 153-8902, Japan \footnote{Present address: International Center
for Materials Nanoarchitectonics,  National Institute for Materials Science,  Tsukuba, Ibaraki 305-0044, Japan} }
\ead{KOMATSU.Hisato@nims.go.jp}
\begin{abstract}
It is difficult to derive the solid-fluid transition theoretically from microscopic models, although this phenomenon itself has been investigated for a long time. We previously constructed an exactly-solvable model with the solid-fluid transition. This model resembles the infinite-range (or mean-field) model in spin systems in some points, hence it can be called a ``mean-field model'' of the solid-fluid transition. In the present paper, we construct a ``mean-field approximation'' of the solid-fluid transition by using the ``mean-field model'' introduced in our previous study, and tries to describe the phase transitions of the three-dimensional Lennard-Jones model as an example. This approximation succeeds in describing the phase diagram which contains three (gas, liquid, and fcc-solid) phase, at least qualitatively.
\end{abstract}

\section{Introduction \label{intro}}
Theoretical understanding of the solid-fluid phase transition from microscopic Hamiltonian is one of the fundamental problems of condensed matter physics and statistical physics. 
Although many numerical studies have been performed to develop the theory of solid-fluid transition\cite{Alder1,Alder2,AK,MS}, most of analytic approaches are based on phenomenological theories, and some do not take into account the translational symmetry breaking explicitly\cite{LJD,MOI,KirMon}. Only a few of them, including the density functional theory (DFT), have successfully described the solid-fluid transition without using phenomenological theories\cite{RY,CA,DA,CL}. 

In the previous study, we considered an exactly-solvable model of two- or three-dimensional classical particle system which shows a solid-fluid transition\cite{KH}. In this model, particles interact with each other only by the sum of cosine potentials. It is a kind of extension of similar researches for one-dimensional system \cite{CF,FP}. We succeeded in making the exactly-solvable model of solid-fluid transition which form triangular, face centered cubic (fcc), body centered cubic (bcc), or simple cubic lattice in the low temperature phase. However, this model has some features different from normal solid-fluid transition. For example, the crystal structure of solid phase of this model depends on the wave numbers of cosine potentials, and does not have more than two crystal structures under a given set of wave numbers. Hence, we should consider how to treat polymorphism, or the possibility to have structural phase transitions of solid phases with different crystals. Furthermore, this model obey the ideal-gas law even in the solid phase, because cosine potentials do not care to which of lattice points the particles come in the solid phase and have no effect on the macroscopic density or volume.
 In our previous study,  we defined the solid phase as the state that the order parameters, or the Fourier components of density corresponding to certain wave numbers, have nonzero values. Hence, this unphysical behavior did not conflict with the definition of the solid phase. However, we should examine whether this transition is really related with normal solid-fluid transition.

In this paper, we study the solid-fluid transition of particle systems with more general two-body isotropic potentials $U( \vector{r})$. The basic idea of the approximation is based on the van der Waals theory. This theory modifies ideal-gas by considering the excluded volume effect and attractive force, and succeeded in describing the liquid-gas transition\cite{Goldenfeld}. Here, the effect of the attractive force is evaluated by a kind of mean-field approximation. Our approximation describes the liquid-gas transition by introducing excluded volume effect and this ``mean-field'' atractive force like the van der Waals theory, and treat the solid-fluid transition by the method which resembles that of the model of our previous study. 

The outline of this paper is as follows: we introduce the idea and the method of the approximation in the former half of section \ref{methodology}, investigate the phase transitions of three-dimensional Lennard-Jones model numerically in the latter half, modify the approximation in section \ref{VK}, and finally summarize these descriptions in section \ref{Discussion}. We also review the van der Waals theory in \ref{vdW_MF}, in order to compare it with our approximation.


\section{Approximation which fix the size of lattice \label{methodology}}

In this section, we introduce the basic method of approximation. This approximation uses Fourier components of the potential corresponding to $\vector{k} = 0$ and the smallest reciprocal lattice vectors of the crystal. Solid-fluid transition is described by the latter Fourier components which is treated by the method similar to the exactly-solvable model considered in our previous study\cite{KH}. Hence, we first review our exactly-solvable model in subsection \ref{Komatsu}, then introduce basic idea of the approximation in \ref{Basic idea}. In the case that the potential diverge sufficiently fast near each particle, we should modify the approximation by considering the excluded volume effect. We explain this modification in detail in subsection \ref{modification}, and apply this approximation to the Lennard-Jones model in \ref{LJ}. 

\subsection{Behavior of the ``mean-field model'' introduced in our previous paper\label{Komatsu}}
In our previous paper\cite{KH}, we considered the $d$-dimensional classical particle model given by
\begin{equation}
 H_{ \left\{ \vector{k}_\alpha \right\} } = \sum _{i} \frac{\vector{p}_i ^2}{2m} - \frac{J}{N} \sum _{\alpha }  \sum _{i,j} \cos \vector{k}_{\alpha } \cdot ( \vector{x}_i - \vector{x}_j ) , 
 \label{HamiltonianMF}
\end{equation}
where $\left\{ \vector{k}_{\alpha } \right\}$ is the set of the smallest reciprocal lattice vectors of the crystal which we consider\cite{KH}. 
This model resembles the one-dimensional exactly-solvable model introduced by earlier research of Carmesin and Fan\cite{CF}, but is different in the point that it lacks short-range repulsive force. This feature enables us to treat our model in two- or higher- dimensions.

The partition function of this model in the limit $N \rightarrow \infty$ is calculated exactly by using saddle-point method, and is given as
\begin{equation}
 Z_{ \left\{ \vector{k}_\alpha \right\} } = \frac{V^N}{\Lambda ^{dN} N!} \left[ \frac{1}{V_{\mathrm{p} } } \max _{r >0 } \left[ \exp \left( - n' \beta J  r^2 \right) \Phi _{\mathrm{p} } (r) \right] \right] ^N ,
 \label{Z2}
\end{equation}
\begin{equation}
\mathrm{where} \left\{
\begin{array}{c}
\displaystyle \Phi _{\mathrm{p} } ( \vector{r} )  \equiv  \int _{ \mathrm{primitive \ cell} } d \vector{x}  \cdot \exp \left( 4 \beta J r \sum _{\alpha } \cos \vector{k} _{\alpha } \cdot \vector{x}  \right) , \\
\displaystyle V_{\mathrm{p} }  \equiv  \int _{\mathrm{primitive \ cell} } d \vector{x}  \cdot 1 . \\
\end{array}
 \right.
 \label{phi_pr}
\end{equation}
The existence of the solid-fluid transition is confirmed by calculating the order parameter $r$. Here, $N, V$,$\beta$, and $n'$  mean the number of particles, the volume, the inverse temperature of the system, and the number of the smallest reciprocal lattice vectors respectively. Some coefficients in (\ref{phi_pr}) and $n'$ are different from those in our previous paper, because we cease to regard the pair of vectors $\pm \vector{k}_{\alpha } $ as one vector in this paper.
The method used in this calculation resembles that of infinite-range models, or mean-field models in spin systems, hence this model can be called a ``mean-field model'' of solid-fluid transition.

The dependence of the partition function given by (\ref{Z2}) on the volume of the system $V$ is the same as that of ideal gas, hence this system obeys the ideal-gas law. This strange property comes from the feature of cosine potentials. They gather particles on particular lattice points, but do not care which one among these points the particles come to. Hence, they have no effect on the macroscopic density or volume.

\subsection{Basic idea \label{Basic idea}}
 From this subsection, we introduce a method to consider the solid-fluid transition of particle systems with two-body isotropic potentials $U( \vector{r})$, i.e. 
\begin{equation}
 H = \sum _{i} \frac{\vector{p}_i ^2}{2m} + \sum _{i,j} U( \vector{r} _i - \vector{r} _j ) . \label{Hamiltonian1}
\end{equation}
First, we regard the system as a box whose length in $x,y$ and $z$-direction are given by $L_x ,L_y$ and $L_z $, respectively, and impose the periodic boundary condition assuming that the decay of the potential $U( \vector{r} )$ in distance is sufficiently rapid.
Using the Fourier transformation of the potential
\begin{equation}
 U ( \vector{r} ) =  \frac{1}{L_x L_y L_z } \sum _{\vector{k}} U _{\vector{k}} e^{i \vector{k} \cdot \vector{r} } \ , \ U _{\vector{k}} = \int _{\Omega } d \vector{r} U ( \vector{r} ) e^{-i \vector{k} \cdot \vector{r} } ,
 \label{U_Fourier}
\end{equation}
the second term of Hamiltonian is expressed as
\begin{eqnarray}
 \sum _{i,j} U ( \vector{r} _i - \vector{r} _j )  & = & \frac{1}{N} \frac{ N}{L_x L_y L_z } \sum _{i,j} \sum _{\vector{k}} U _{\vector{k}} \exp i \vector{k} \cdot \left( \vector{r} _i - \vector{r} _j \right)  \nonumber \\
  & = & \frac{1}{N} \rho \sum _{i,j} \sum _{\vector{k}} U _{\vector{k}} \cos \vector{k} \cdot \left( \vector{r} _i - \vector{r} _j \right)  .
  \label{Pfrr}
\end{eqnarray}
The cosine potentials in the right hand side of (\ref{Pfrr}) can be transformed into one-body potentials with auxiliary variables by Hubbard-Stratonovich transformation, and these auxiliary variables coincide with the Fourier components of density as in the infinite-range XY model or our previous study\cite{KH}. However, considering all wave numbers is too difficult, hence we take only a few wave numbers important in phase transitions into account. Namely, we ignore all wave numbers except $\vector{k} = 0$ and the smallest reciprocal lattice vectors(RLVs) of the crystal;
\begin{eqnarray}
  \sum _{i,j} U ( \vector{r} _i - \vector{r} _j ) & \sim & \frac{1}{N}  \rho \sum _{i,j} \sum _{\vector{k}:\mathrm{0 \ or \ smallest \ RLV}} U _{\vector{k}} \cos \vector{k} \cdot \left( \vector{r} _i - \vector{r} _j \right) \nonumber \\
  & = &  \rho U_0 N + \frac{1}{N} \rho \sum _{\vector{K}:\mathrm{ smallest \ RLV}} U _{K} \cos \vector{K} \cdot \left( \vector{r} _i - \vector{r} _j \right) .
 \label{Approx1}
\end{eqnarray}
In the second line of (\ref{Approx1}), we used the fact that $U _{\vector{K}}$ only depends on $K$, or the absolute value of $\vector{K}$. Note that the values of $\vector{k}$ are limited to the form $( 2 \pi n_x / L_x , 2 \pi n_y / L_y , 2 \pi n_z / L_z )$ where $n_i$ are integers. Hence, we change the value of $L_i$ slightly in order to let $\vector{K}$ fulfill this condition assuming that this manipulation scarcely affect the behavior of the system.
Comparing the cosine potential of (\ref{Approx1}) with that of (\ref{HamiltonianMF}), $- \rho U _{K} $ can be regarded as the effective coupling constant in (\ref{Approx1}). The reason why the effective coupling constant includes the density $\rho$ is the existence of the coefficient $\frac{1}{L_x L_y L_z } $ in the Fourier expansion (\ref{U_Fourier}). As we saw in (\ref{Pfrr}), this coefficient generates $\rho$.
We calculate an approximate partition function by integrating (\ref{Approx1}). Assuming that $U _{K}$ is negative, the right hand side of (\ref{Approx1}) has the same form as (\ref{HamiltonianMF}), except the constant term, $\rho U_0 N$. Hence the result of integration also has the same form as (\ref{Z2});
\begin{equation}
 Z = \exp \left[ C.E. _{ (V )} \log \left[ \frac{V^N}{N! \Lambda ^{dN}} \max _{r} \left\{ e^{-\beta \rho U_0} e^{-n' \beta \rho | U_{K} | r^2} \frac{\Phi _p ( 4 \beta \rho | U_{K} | r )}{V_p} \right\} ^N \right] \right] ,
 \label{Z3}
\end{equation}
\begin{equation}
 \mathrm{where} \ \Phi _p ( a ) =  \int _{\mathrm{primitive \ cell} } d \vector{x}  \cdot \exp \left\{ \sum _{\vector{K}:\mathrm{ smallest \ RLV}} \frac{a}{2} \cos \vector{K} \cdot \vector{x} \right\} .
\end{equation}
Here, $ C.E. _{ (V )} $, which means the convex envelope about $V$, is introduced in order to keep the convexity of free energy. If $U _{K}$ has the positive value, solid phase does not appear because the cosine potentials prevent the system from making the order. The coefficient $e^{-\beta \rho U_0}$ in (\ref{Z3}) is the same form as the contribution of ``attractive force'' in the van der Waals theory, and hence causes the liquid-gas transition. This affinity comes from the fact that the van der Waals theory also estimates the effect of ``attractive force'' by its mean value over the space, $U_0$. The main differences between our approximation and the van der Waals theory is that we also take $U _{K}$ into account. We discuss the relation between these two theories in more detail in the appendix.
After the Legendre transformation with respect to pressure $p$, we obtain the expression related with the Gibbs free energy;
\begin{eqnarray}
 Z_G & = &  \max _{\rho } \left( e^{-\frac{\beta N p}{\rho} } Z \right) \nonumber \\
& = & \max _{r , \rho } \left[ \frac{ V^N }{N! \Lambda ^{dN}}   \left\{  e^{-\beta \rho U_0 -n' \beta \rho | U_{K} | r^2 -\frac{\beta p}{\rho}} \frac{\Phi _p ( 4 \beta \rho | U_{K} | r )}{V_p} \right\} ^N \right] \label{ZG} .
\end{eqnarray}
Solid or fluid phases are discerned by whether the value of argument $r$ of the maximum in (\ref{ZG}) has nonzero value or not.
When the system has a possibility to form several different crystal structures, we should investigate all candidates of the structures in (\ref{ZG}), and choose one which has the largest $Z_G$, or the smallest Gibbs free energy. 
The phase transition between two fluid phase, or liquid and gas phase, is expressed as the discontinuity of argument $\rho$ of the maximum in (\ref{ZG}), but we should modify this equation considering the excluded volume effect in order to describe the liquid-gas transition. This modification is explained in the next subsection.

\subsection{Consideration of the excluded volume effect \label{modification} }
For models whose potentials diverge sufficiently fast at $\vector{r} = 0$ such as the Lennard-Jones model, the Fourier transformation of the potentials also diverge. Hence we need to exclude the neighborhood of $\vector{r} = 0$ from the calculation of the Fourier transformation, and modify the approximation of subsection \ref{Basic idea} by reflecting the contribution of the neighborhood in other way.

We assume that the potential is attractive outside a certain radius, and repulsive inside it. In this case, particles rarely come close to each other.
 Hence, we divide the space into the Wigner-Seits cells of the lattice, and take the effect of repulsive force into consideration by assuming that each Wigner-Seits cell can have at most one particle. Taking into account the fact that the number of Wigner-Seits cells is given by $N_c \equiv V/V_{\mathrm{p}} $, the coefficient $\frac{N_c ^N}{N!} = \frac{(V/V_p)^N}{N!}$ in the partition function, which corresponds to the number of ways to distribute particles to the Wigner-Seits cells under the condition that the number of particles existing in each Wigner-Seits cell is not limited, is  replaced with
\begin{equation}
\displaystyle  _{N_c} C_N = _{(V/V_p)} C_N  \sim \frac{\left( \frac{V}{V_p} \right)^{\frac{V}{V_p} } }{ \left( \frac{V}{V_p} - N \right)^{\frac{V}{V_p} - N} \cdot N^N } =  \left\{ \left( 1- \rho V_p \right)^{ -\left( 1- \rho V_p  \right) / \rho V_p} \left( \rho V_p \right)^{-1} \right\} ^N , \label{dist1}
\end{equation}
corresponding to that under the condition that each Wigner-Seits cell accept at most one particle. Here, we used Stirling's formula in (\ref{dist1}). According to the above assumptions, we modify (\ref{ZG}) by the following rules;
\begin{itemize}
 \item We consider only the attractive part of the potential for Fourier transformation $U_0 $ and $U_{K} $.
 \item We replace the coefficient $\frac{(V/V_p)^N}{N!}$ in the partition function with $\left\{ \left( 1- \rho V_p \right)^{ -\left( 1- \rho V_p  \right) / \rho V_p} \left( \rho V_p \right)^{-1} \right\} ^N $.
\end{itemize}
$Z_G$ is modified under this rule as
\begin{eqnarray}
 Z_G & = & \left[ \frac{1}{\Lambda ^{dN}} \max _{r , \rho } \left\{ \left( 1- \rho V_p \right)^{ -\left( 1- \rho V_p  \right) / \rho V_p} \left( \rho V_p \right)^{-1} \right. \right. \nonumber \\
 & & \left. \left. \cdot e^{-\beta \rho U_{a 0} -n' \beta \rho | U_{a K} | r^2 -\frac{\beta p}{\rho}} \Phi _p ( 4 \beta \rho | U_{a K} | r ) \right\} ^N \right] ,
\label{ZG2} 
\end{eqnarray}
where $U_a$ is the attractive part of $U$.
Our approximation resembles that of the DFT of early ages because it uses a ``mean-field approximation'' to the Fourier components of density corresponding to reciprocal lattice vectors\cite{RY}. Hence, mathematical aspect of (\ref{ZG2}) is similar to that of the early DFT. However, our approximation starts from an exactly-solvable model, and as a result, a bare potential appears in (\ref{ZG2}). This point is large difference from DFT which uses direct correlation function, or a kind of effective potentials\cite{ RY,CA,DA}.

\subsection{The phase transitions of the three-dimensional Lennard-Jones model \label{LJ} }
\begin{figure}[!hbp]
\begin{center}
\includegraphics[width = 8.0cm]{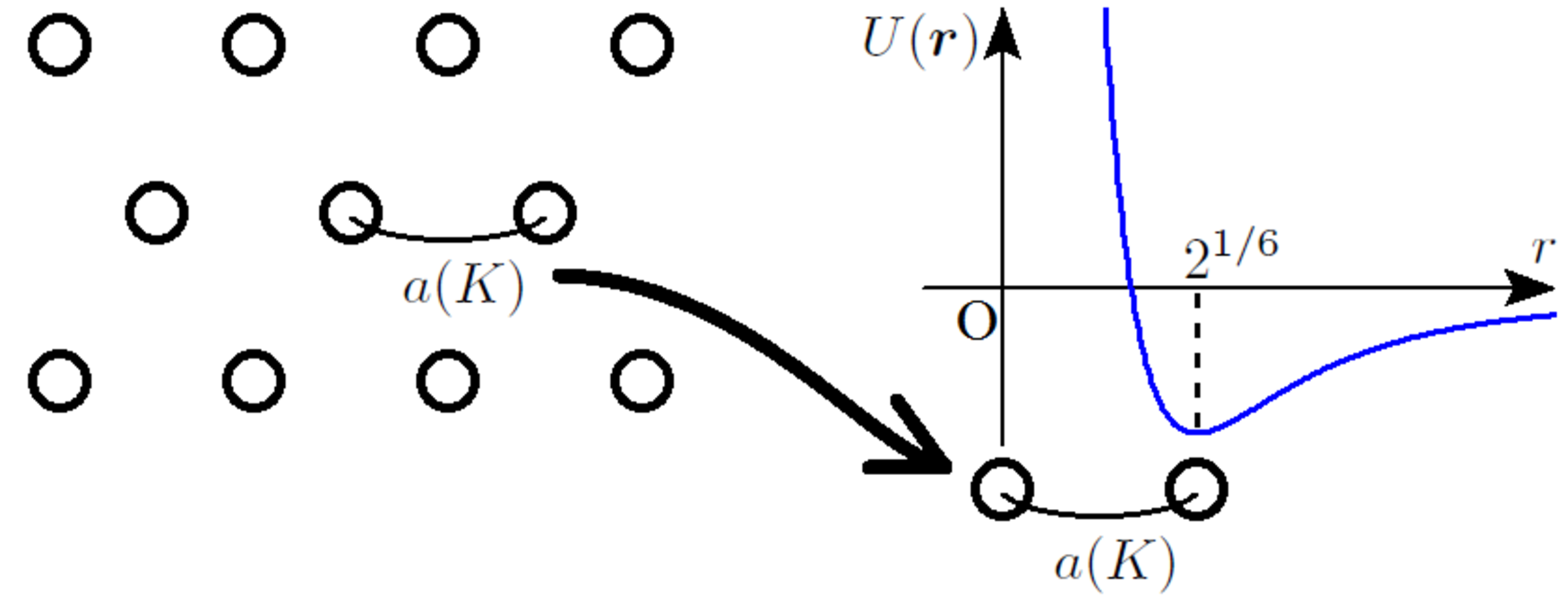}
\caption{Perfect crystal (left) and the minimum of Lennard-Jones potential (right) : We define $a(K)$ as the distance between adjacent particles of perfect crystal. This value can be expressed as the function of $K$ because the size of the perfect crystal is determined when the size of the smallest reciprocal lattice is given. In section \ref{methodology}, we assume that $a(K)$ coincides with the argument of minimum of Lennard-Jones potential, $2^{1/6}$. The value of $K$ can be calculated by using this relation. }
\label{aK1}
\end{center}
\end{figure}

The Lennard-Jones model is frequently used as the model of classical simple liquids. Its potential is given by
\begin{equation}
 U( \vector{r} ) = 2 \epsilon \left\{ \left( \frac{\sigma }{r} \right) ^{12} - \left( \frac{\sigma }{r} \right) ^6 \right\} .
\end{equation}
 Although this potential includes two constants $\epsilon$ and $\sigma $, because physical quantities are nondimensionalized by putting $\vector{r} ^{\ast } = \vector{r} / \sigma ,  \beta^{\ast } = \beta \epsilon$ and $p^{\ast } = p \sigma^3 / \epsilon$, these constants are regarded as $1$ without losing generality. In order to use (\ref{ZG2}), we should determine the magnitude of the smallest reciprocal lattice vectors $K$. In this section, we determine the value of $K$ so that the distance between adjacent particles in the perfect crystal $a(K)$ coincides with the radius which gives the potential minimum as shown in figure \ref{aK1}, 
\begin{equation}
 a(K ) = 2^{1/6} .
 \label{K1}
\end{equation}

\begin{figure}[!hbp]
\begin{center}
\includegraphics[width = 11.0cm]{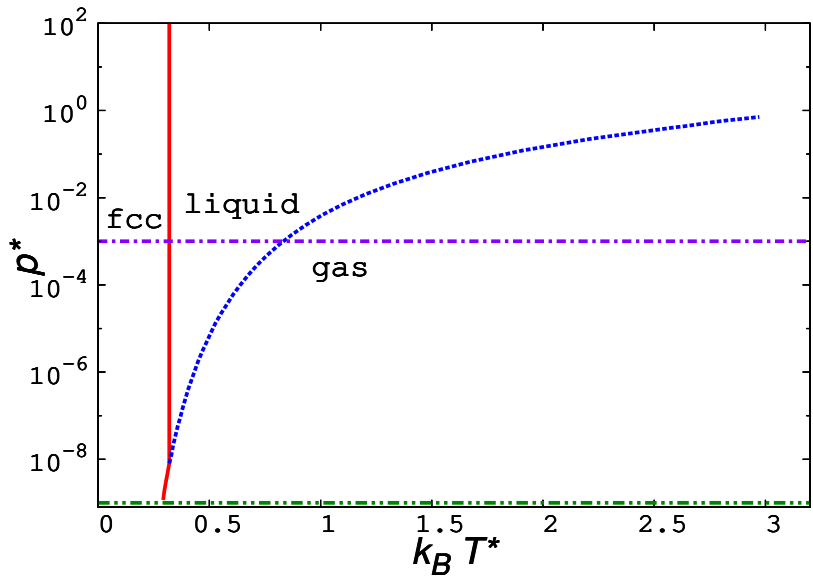}
\caption{The phase diagram of Lennard-Jones model obtained by the approximation explained in section \ref{methodology} : The red solid line is the fcc-fluid transition temperature, and the blue dashed one is the liquid-gas transition temperature. We also investigated the behavior of the order parameter along purple chain and green chain double-dashed lines and described it at figure \ref{Frr}. }
\label{LJ-PD}
\end{center}
\end{figure}

\begin{figure}[!hbp]
\begin{center}
\includegraphics[width = 11.0cm]{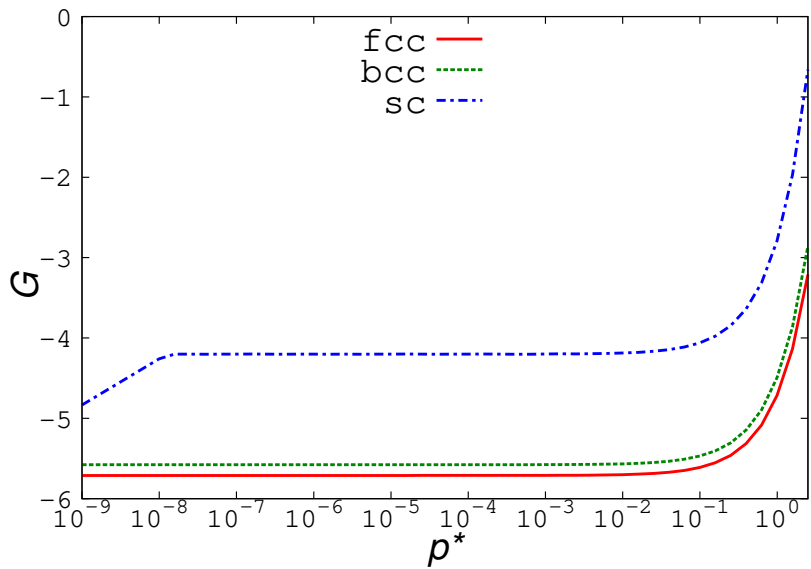}
\caption{The Gibbs free energy calculated by the approximation explained in section \ref{methodology} at $k_B T^* = 0.25$ : The red solid line is the Gibbs free energy of fcc lattice, the green dashed one is that of bcc lattice, and the blue chain one is that of simple cubic lattice. We ignored the term derived from kinetic energy because it does not affect this comparison. Winding of the free energy of simple cubic lattice at low pressure is caused by transition into the fluid phase. }
\label{Free}
\end{center}
\end{figure}

\begin{figure}[!hbp]
\begin{center}
\includegraphics[width = 11.0cm]{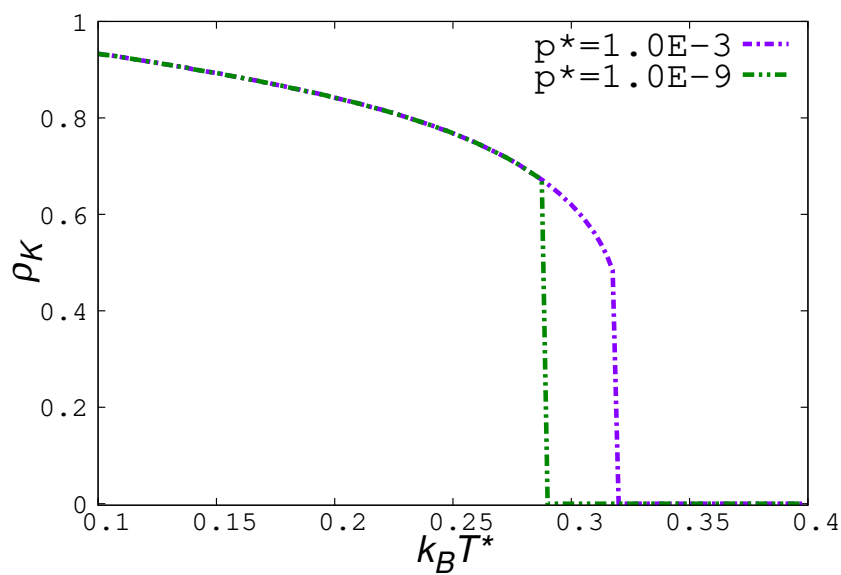}
\caption{Temperature dependence of the order parameter of the Lennard-Jones model obtained by the approximation of section \ref{methodology} : The purple chain line stands for the data of $p^* =10^{-3}$, and the green chain double-dashed line stands for $p^* =10^{-9}$.}
\label{Frr}
\end{center}
\end{figure}

 Calculating $K$ from (\ref{K1}) using elementary geometry, we obtain $K=3^{1/2} 2^{1/3} \pi \simeq 6.856$ for fcc and bcc, and $K=2^{5/6} \pi \simeq 5.598$ for simple cubic lattice.
As a result of numerical investigation of (\ref{ZG2}) on fcc, bcc, and simple cubic lattices, we have the $p-T$ phase diagram of the three-dimensional Lennard-Jones model shown in figure \ref{LJ-PD}. Note that bcc and simple cubic lattices do not appear in the phase diagram because their free energies are higher than that of fcc lattice as shown in figure \ref{Free}. This fact corresponds to the results of simulation \cite{AK,MS}. Although we succeed in describing the phase transition among gas, liquid, and fcc-solid phase, this phase diagram has some problems. For example, the melting temperature does not increase when the pressure is higher than that of triple point. This is because the value of $\rho U_{aK}$, which corresponds to the coupling constant of the ``mean-field model'', has an upper limit, and as a result the temperature where the solid phase can exist is limited. In fact, according to the numerical data, the value of density $\rho$ is almost equal to the upper limit in the solid and liquid phase at the transition temperature. Furthermore, as a result of the upper limit of $\rho U_{aK}$, the temperature dependence of the order parameter of the solid phase described in figure \ref{Frr} resembles that of the ``mean-field model''. 

\clearpage
\section{Modification including change of the size of lattices \label{VK}}

As we saw in section \ref{methodology}, the approximation we introduced had some problems such as the upper limit of transition temperature. In this section, we do not fix the value of $K$ in order to solve these problems. Under this modification, the size of lattice changes depending on $K$, hence we have to improve the details of approximation. We first discuss a difficult point towards this modification in subsection \ref{Problems}, then introduce an improved method for estimating $U_K$ in \ref{Jeff}, and then draw the phase diagram of the Lennard-Jones model in \ref{LJvK}. 

\subsection{Problems which occur if the value of $K$ is not fixed\label{Problems}}
The problems we referred to in subsection \ref{LJ} result from the approximation in which we fix the magnitude of the smallest reciprocal lattice vectors $K$ and corresponding Fourier component of the potential $U_{aK}$. Hence these problems seem to be resolved if we stop fixing $K$, but this trial does not succeed easily. This subsection discuss the difficulty in changing the value of $K$.
First of all, the size of the Wigner-Seits cell is proportional to $K^{-1}$. As a result, if there is a possibility that $K$ have the large value, it is necessary to consider the case in which the distance between adjacent particles become sufficiently small for the repulsive force to be dominant. Therefore, the cut-off of the integration of Fourier transformation should be shifted 
 to more appropriate value such as 
\begin{equation}
\sigma (K) \equiv \sqrt[d]{ \frac{ \Gamma \left( \frac{d}{2} + 1 \right) }{ \pi^{\frac{d}{2} } } V_p }  ,
\label{rmin}
\end{equation}
or the radius of the ball whose volume $V_p$ coincides with the Wigner-Seits cell of the lattice (see figure \ref{sigmaK1}). However, 
under this modification, inaccuracy of the estimation of the energy grows larger with the increase of $K$, if we calculate $U_K$ by its original definition given by (\ref{U_Fourier}).
 Hence, we should also modify the method of estimating $U_K$. In this section, we modify the method of estimating $U_K$ and apply the modified approximation to Lennard-Jones model, whereas the form of (\ref{ZG2}) itself is almost unchanged except the kind of variables over which we should seek the maximum. 

\subsection{Estimation of $U_K$ \label{Jeff}}
In this subsection, we explain how to modify the estimation of $U_K$. It is done by the method different from its original definition (\ref{U_Fourier}). 
With changing the value of $K$, the energy of the system is incorrectly estimated in the solid state, under the approximation of the previous section. In order to estimate the energy of the system correctly in the solid state, especially in the perfect crystal, we first divide the energy of the perfect crystal into the contribution of $U_0$ and $U_K$.

We let the distance between adjacent particles in the perfect crystal state with given value of $K$ be $a(K)$, and the number of Wigner-Seits cells surrounding one cell be $n_n$. Using these values, the sum of the potentials between one particle and adjacent particles in the perfect crystal, $E_0$, is given as follows:
\begin{equation}
E_0 = n_n U \left( a(K) \right) .
\end{equation}
Furthermore, we put the product of $n_n$ and the average value of potential over adjacent cells as $A_0$:
\begin{equation}
A_0 = n_n \cdot \frac{\displaystyle  \int _{\mathrm{ adjacent \ cells } } U( \vector{r} ) d\vector{r} }{\displaystyle  \int _{\mathrm{ adjacent \ cells } } d\vector{r} } , 
\label{A00}
\end{equation}
 then interpret it as a part of the contribution of $U_0$, and regard the subtraction $(E_0 - A_0 )$ as the contribution of $U_K$. 

\begin{figure}[!hbp]
\begin{center}
\includegraphics[width = 8.0cm]{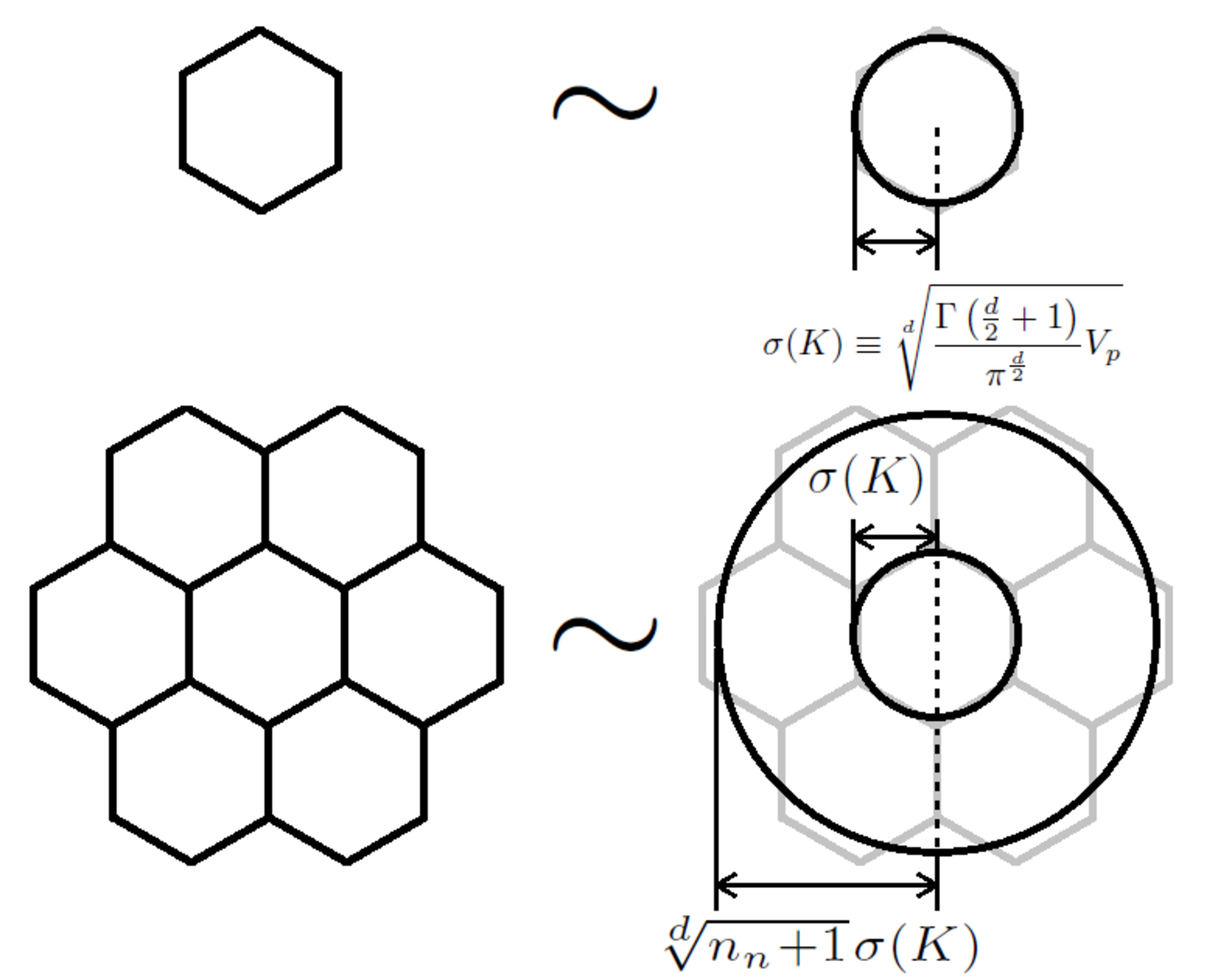}
\caption{Schematic figure for the definition of $\sigma_K$ and range of the integration in (\ref{A0}) : We define $\sigma_K$ as the radius of the ball with the same volume as the Wigner-Seits cell of the lattice. Similarly, we approximate the sum of one cell and adjacent cells as the ball with the same volume as it. Hence, the sum of adjacent cells is approximated as a spherical shell with inner radius $\sigma(K)$ and outer radius $\sqrt[3]{ n_n +1} \sigma (K)$ when we calculate $A_0$ by (\ref{A0}). }
\label{sigmaK1}
\end{center}
\end{figure}

Here, we approximate $A_0$ as the average over spherical shell with the same volume as the sum of adjacent Wigner-Seits cells as illustrated in figure \ref{sigmaK1} for simplicity.
 Namely, we calculate $A_0$ as
\begin{equation}
A_0 = n_n \cdot \frac{\displaystyle  \int _{\sigma (K)} ^{ \sqrt[d]{ n_n +1} \sigma (K)} U(r) r^2 dr }{\displaystyle  \int _{\sigma (K)} ^{ \sqrt[d]{ n_n +1} \sigma (K)} r^2 dr} . 
\label{A0}
\end{equation}
 Here, we assume that the effect of farther cells to $U_K$ is negligible. 

Under the approximation that express the potential as (\ref{Approx1}), the energy difference between the perfect crystal and the fluid state where particles is placed randomly is calculated as
\begin{equation}
 \frac{1}{N} \rho \sum _{i,j} \sum _{\vector{K}:\mathrm{ the \ smallest \ RLV}} U _K = N n' \rho U_K .
\end{equation}
We assume that this difference per one particle is equal to the difference between $E_0$ and $A_0$, namely,
\begin{equation}
 n' \rho U_K = E_0 - A_0.
\end{equation}
As a result, the effective coupling constant in the perfect crystal state is expressed as:
\begin{equation}
\rho U_K = \frac{E_0 - A_0}{n'} . \label{UK_perfect}
\end{equation}
If there are cells which do not have a particle, the probability that each cell has a particle is given by 
\begin{equation}
\frac{N}{N_c} = \frac{N}{(V/ V_{\mathrm{p}} )} = \rho V_{\mathrm{p}} .
\end{equation}
We evaluate the effective coupling constant of this case by the product of the right hand side of (\ref{UK_perfect}) and this probability:
\begin{equation}
\rho U_K = \rho V_{\mathrm{p}} \frac{E_0 - A_0}{n'} .
\label{UK_VK}
\end{equation}

Although this evaluation is different from the original definition of $U_K$ given by (\ref{U_Fourier}), inaccurate estimation of energy is avoided by adopting it. Note that $U_0 $ is estimated by Fourier transformation with cut-off as before. Here, the cut-off is introduced at the radius of the ball whose value is the same as the Wigner-Seits cell of the supposed crystal, $\sigma (K)$. Namely, $U_0$ is estimated as:
\begin{equation}
U_0 = \int _{r \geq \sigma (K)} d \vector{x} U (\vector{x} ) = 4 \pi  \int _{\sigma (K)} ^{\infty} r^2 dr U (r) ,
\label{U0_VK}
\end{equation}
The form of the partition function itself is almost unchanged from (\ref{ZG2}) except the point that we should take maximum over $K$, in addition to $r$ and $\rho$.
\begin{eqnarray}
 Z_G & = & \left[ \frac{1}{\Lambda ^{dN}} \max _{r , \rho , K } \left\{ \left( 1- \rho V_p \right)^{ -\left( 1- \rho V_p  \right) / \rho V_p} \left( \rho V_p \right)^{-1} \right. \right. \nonumber \\
 & & \left. \left. \cdot e^{-\beta \rho U_{ 0} -n' \beta \rho  \left| \min (U_K , 0) \right| r^2 -\frac{\beta p}{\rho}} \cdot \Phi _p \left( 4 \beta \rho  \left| \min (U_K , 0) \right| r \right) \right\} ^N \right] ,
\label{ZG3} 
\end{eqnarray}
Here, we assume that $U_K$ does not affect the behavior of the system if it is positive.

At the end of this subsection, we discuss the dependence of the partition function and the free energy in the fluid phase on the crystal structure. 
The relation
\begin{equation}
 \Psi _{\mathrm{p}} ( 0 ) =  V_p ,
\end{equation}
holds because of (\ref{phi_pr}), or the definition of $\Psi _{\mathrm{p}} $. Hence, in the fluid state, (\ref{ZG3}) is transformed as
\begin{equation}
 Z_G = \left[ \frac{1}{\Lambda ^{dN}} \max _{ \rho ,K } \left\{ \left( 1- \rho V_p \right)^{ -\left( 1- \rho V_p  \right) / \rho V_p} \rho ^{-1}  e^{-\beta \rho U_{0} -\frac{\beta p}{\rho}} \right\} ^N \right] . \label{ZGfluVK}
\end{equation}
Remarking that the $U_0$ depend on $K$ and assumed crystal structure only by $V_p$ as we can see from (\ref{U0_VK}) and (\ref{rmin}), $Z_G$ given by (\ref{ZGfluVK}) also. Letting the value of $V_p$ when $K=1$ be $V_{p1} $, the value of $V_p$ for given $K$ is expressed as 
\begin{equation}
V_p =  \frac{V_{p1} }{K^d},
\end{equation}
because $V_p$ is proportional to the inverse-$d$th power of $K$.
Change of crystal structure affects the right hand side of (\ref{ZGfluVK}) by the value of $V_{p1} $. However, even if $V_{p1} $ is changed, we can keep the value of $V_p$ unchanged by changing the value of $K$. Hence, the maximum about $K$ appearing in the right hand side of (\ref{ZGfluVK}) has the same value: 
\begin{equation}
 Z_G = \left[ \frac{1}{\Lambda ^{dN}} \max _{ \rho } \left\{ \left( 1- \rho {V_{\mathrm{max}}} (\rho ) \right)^{ -\left( 1- \rho {V_{\mathrm{max}}} (\rho )  \right) / \rho {V_{\mathrm{max}}} (\rho ) } \rho ^{-1}  e^{-\beta \rho U_{0} -\frac{\beta p}{\rho}} \right\} ^N \right] , \label{ZGfluVK2}
\end{equation}
for any crystal structure. Here, $V_{\mathrm{max}} (\rho )$ is the value of $V_p$ which gives the maximum of the inside of the curly bracket of (\ref{ZGfluVK}). This value is realized when $K$ is expressed as: 
\begin{equation}
K =  \sqrt[d]{ \frac{V_{p1} }{V_{\mathrm{max}} (\rho ) } } .
\end{equation}
From this discussion, whichever crystal structure we assume, the partition function and the free energy of fluid phase do not change. This feature is not seen in the method we treated in section \ref{methodology} which fixed the value of $K$.

\subsection{Phase diagram of Lennard-Jones model \label{LJvK}}

\begin{figure}[!hbp]
\begin{center}
\includegraphics[width = 11.0cm]{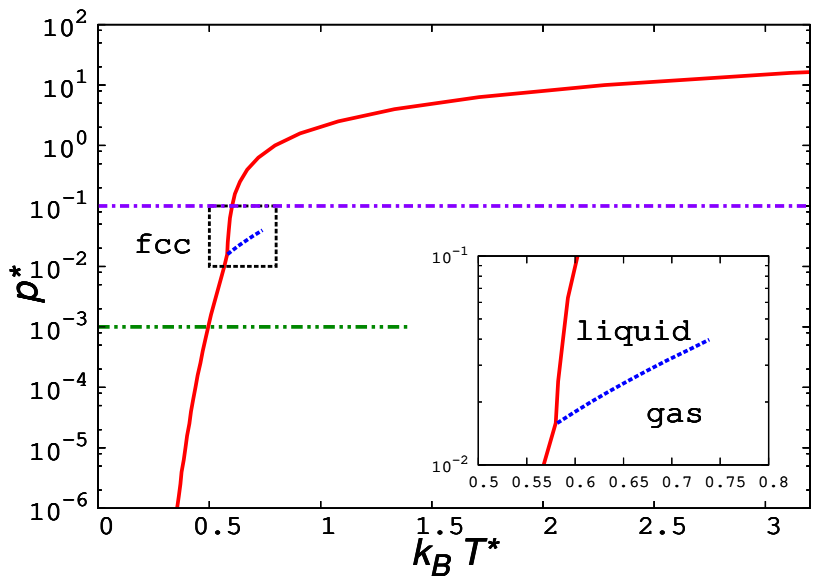}
\caption{The phase diagram of Lennard-Jones model obtained by the approximation of section \ref{VK} : The domain inside the black frame is expanded in the inset. The result of the approximation coincides with that of simulation at least qualitatively\cite{AK,MS}. We also investigated the behavior of the order parameter along purple chain and green chain double-dashed line and described it at figure \ref{Frr_VK} like section \ref{methodology}.}
\label{LJ-PD_VK}
\end{center}
\end{figure}

Evaluating the maximum of (\ref{ZG3}) numerically, we investigate three-dimensional Lennard-Jones model and show the phase transition with the phase diagram in figure \ref{LJ-PD_VK}. Like section \ref{methodology}, we evaluate the free energy of bcc, fcc and simple cubic lattices as the candidates of crystal structure of the solid phase. As a result, fcc lattice turns out to have the lowest free energy and appears as the solid phase. Figure \ref{LJ-PD_VK} shows the phase diagram obtained by the approximation, which resembles that obtained by simulation\cite{AK,MS}, especially in the point that the upper limit of melting temperature does not exist as in the fixed-$K$ case in the previous section. 

\begin{figure}[!hbp]
\begin{center}
\includegraphics[width = 11.0cm]{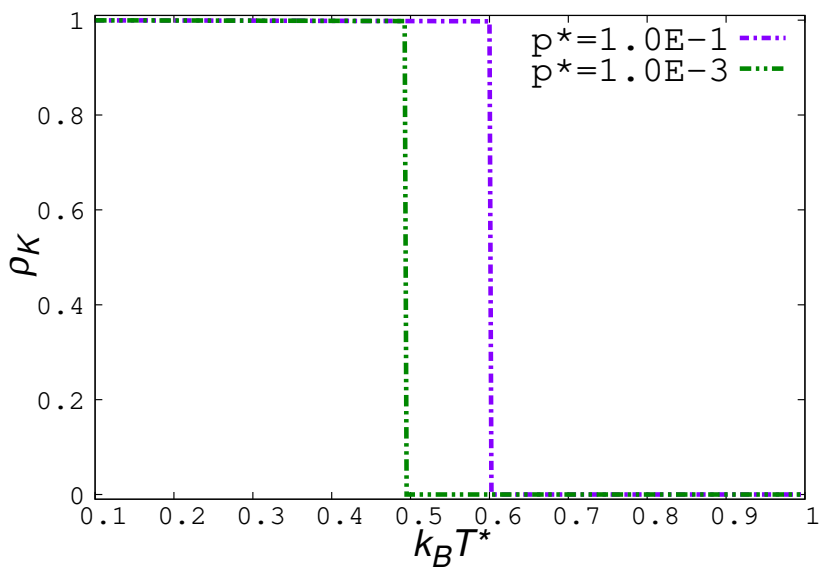}
\caption{Temperature dependence of the order parameter of Lennard-Jones model obtained by the approximation of section \ref{VK} : The purple chain line stands for the data for $p^* =10^{-1} ,$ and the green chain double-dashed line stands for $p^* =10^{-3} ,$ respectively.}
\label{Frr_VK}
\end{center}
\end{figure}

Another difference between the result of this approximation and previous section is the behavior of order parameter. The order parameter obtained by this approximation is almost 1 in the solid phase regardless of the temperature like figure \ref{Frr_VK}. This is because the fluid phase corresponding to other values of $K$ than the solid phase is more stable than the solid phase with smaller value of order parameter in the high temperature region. 

\clearpage
\section{Summary \label{Discussion}}
We introduced in this paper an approximation for classical particle systems which describe the phase transition among gas, liquid and solid phases, and investigated the Lennard-Jones model as an example. This approximation uses Fourier components of the potential corresponding to $\vector{k} = 0$ and the smallest reciprocal lattice vectors of the crystal. The latter Fourier components are important in stabilizing solid phase and treated by the same method as the exactly-solvable model considered in our previous paper. The former Fourier component and modification of partition function reflecting the excluded volume effect cause liquid-gas transition by the mechanism similar to the van der Waals theory.
 The model of our previous study can be called as a ``mean-field model'' of solid-fluid transition, and the van der Waals theory has an aspect of mean-field approximation for liquid-gas transition. Hence, this approximation should be a kind of ``mean-field approximation''.

Although this approximation ignores many Fourier components of potential and also uses oversimplified method to consider the excluded volume effect, it is notable that such rough approximation succeeded in describing the phase transitions between three phases of Lennard-Jones model without changing the form of partition function. It is also interesting that the strange model of our previous study which obey the ideal-gas law even in the solid phase can be utilized to the description of normal solid-fluid transition.

Comparing this approximation with the DFT, one big difference is that this approximation use bare potential, whereas the DFT uses effective one, as we have discussed at the end of subsection \ref{modification}. This means that our method do not need to use approximation specific to liquid theory such as Percus-Yevick approximation, because we do not have to estimate the effective potential. 

We have room for improvement in our approximation, because it has problems such as  many arbitrarinesses in methodology. Hence, further study should be needed. 
 
\section*{Acknowledgements}
The author thanks to Koji Hukushima, Masamichi J. Miyama, Masamichi Nishino, and Yoshihiko Nonomura for helpful discussions. 

\appendix

\section{The van der Waals theory interpreted as a mean-field theory \label{vdW_MF}}
The van der Waals theory is a widely known theory of real gas which can treat even the phase transition into liquid. From a standpoint of modern statistical physics, this theory can be interpreted as a mean-field theory of liquid-gas transition. In this appendix, we review van der Waals theory focusing on the aspect of a mean-field theory and the difference between the approximation introduced in this paper.

We start from the classical particle system given by (\ref{Hamiltonian1}).
The second term of this Hamiltonian is transformed as
\begin{equation}
\sum _{i \neq j} U( \vector{r} _i - \vector{r} _j ) = \sum _i \sum _{j:j\neq i} U( \vector{r} _i - \vector{r} _j ) .
\label{U0_vdW_0}
\end{equation}
We approximate each summation over $j$ as integration over the space:
\begin{equation}
\sum _{j:j\neq i} U( \vector{r} _i - \vector{r} _j ) \sim \left( \frac{V}{N} \right)^{-1} \int _{r \geq r_0} d \vector{r} U ( \vector{r} ) = \rho \int _{r \geq r_0} d \vector{r} U ( \vector{r} ) .
\label{U0_vdW_1}
\end{equation}
Here, the integration is divided by the weight, or the volume per one particle $\frac{V}{N}$. We skip the detailed method of estimating the cut-off of integration $r_0$ in this appendix.
 This approximation replace the summation of potential with its mean-value over the space. Hence it can be called as a kind of mean-field approximations. 
The integration in (\ref{U0_vdW_1}) coincides with the Fourier component $U_0$ except the existence of cut-off, hence (\ref{U0_vdW_0}) is expressed as
\begin{equation}
\sum _{i \neq j} U( \vector{r} _i - \vector{r} _j ) \sim \sum _i \rho U_0 =  \rho U_0 N .
\label{U0_vdW_2}
\end{equation}
This term has the same form as the first term of the right hand side of (\ref{Approx1}). Hence, (\ref{Approx1}) coincides with (\ref{U0_vdW_2}) if we ignore $U_K$, or the Fourier components of potential corresponding to the smallest reciprocal lattice vectors. Using (\ref{U0_vdW_2}), partition function of the system is given as
\begin{equation}
 Z = \exp \left\{ C.E. _{ (V )} \log \left( \frac{V^N}{N! \Lambda ^{dN}}  e^{-\beta N \rho U_0} \right) \right\} .
 \label{ZvdW0}
\end{equation}
In the van der Waals theory, we express the excluded volume effect by reducing the volume of region where each particle can move. Considering this effect, the partition function is modified as
\begin{equation}
 Z = \exp \left[ C.E. _{ (V )} \log \left\{ \frac{(V-Nb)^N}{N! \Lambda ^{dN}}  e^{-\beta N \rho U_0} \right\} \right] .
 \label{ZvdW}
\end{equation}
Here, $b$ is the excluded volume per one particle. Using (\ref{ZvdW}), the free energy is expressed as follows:
\begin{eqnarray}
F & = & - \frac{1}{\beta} \log Z = - \frac{1}{\beta}  C.E. _{ (V )} \log \left\{ \frac{(V-Nb)^N}{N! \Lambda ^{dN}}  e^{-\beta N \rho U_0} \right\} \nonumber \\
& = & - C.E. _{ (V )} \left[ \frac{1}{\beta} \log \left\{ \frac{(V-Nb)^N}{N! \Lambda ^{dN} } \right\} - N \rho U_0 \right] .
\label{FvdW}
\end{eqnarray}
Differentiating both sides of (\ref{FvdW}), we obtain the equation of state: 
\begin{eqnarray}
p = -\left( \frac{\partial F}{\partial V} \right) _{T} & = & M.C. _{(V)} \left\{ \frac{N}{\beta (V-Nb)} + \frac{N^2 U_0}{V^2} \right\} \nonumber \\
& = & M.C. _{(V)} \left\{ \frac{N k_B T}{(V-Nb)} + \frac{N^2 U_0}{V^2} \right\} .
\label{vdWeq_1}
\end{eqnarray}
Here, $ M.C. _{ (V )} $ means the function modified by so called Maxwell equal area rule. In the case that the effect of potential is mainly attractive, $U_0$ has negative value. Hence, (\ref{vdWeq_1}) is also expressed as 
\begin{equation}
p = M.C. _{(V)} \left\{ \frac{N k_B T}{(V-Nb)} - \frac{N^2 | U_0 | }{V^2} \right\} .
\label{vdWeq_2}
\end{equation}
This is the well-known form of van der Waals equation.  In usual cases, this mean-field-like evaluation of attractive force is no more than an approximation, but it becomes exact in systems with some kinds of long range interactions\cite{KUH,LP}.

Comparing the approximation of this paper and the van der Waals theory discussed above, the difference between these two theories comes from the existence of $U_K$ and the method to estimate the excluded volume effect. Nonuniform Fourier component $U_K$ is important only when we consider the solid phase, hence both theories are essentially alike in the fluid phase.
In order to compare the partition functions of both theories concretely, we transform (\ref{ZvdW}) using the Legendre transformation about pressure $p$;
\begin{eqnarray}
 Z_G & = &  \max _{\rho } \left( e^{-\frac{\beta N p}{\rho} } Z \right) = \max _{ \rho } \left\{ \frac{(V-Nb)^N}{N! \Lambda ^{dN}} e^{-\beta N \rho U_0 -\frac{\beta N p}{\rho}}  \right\} \nonumber \\
& = & \max _{ \rho } \left\{ \frac{V^N }{N! \Lambda ^{dN}} \cdot (1-\rho b)^N e^{-\beta N \rho U_0 -\frac{\beta N p}{\rho}}  \right\} \nonumber \\
& \sim &  \frac{1 }{\Lambda ^{dN}} \max _{ \rho } \left\{ (1-\rho b) \rho^{-1} e^{1-\beta  \rho U_0 -\frac{\beta p}{\rho}}  \right\} ^N \label{ZGvdW} .
\end{eqnarray}
The last line of (\ref{ZGvdW}) is obtained by using Stirling's formula.

As we discussed in section \ref{VK}, the partition function is given by (\ref{ZGfluVK2}) in the fluid phase under the approximation of this paper. In the condition that $\rho {V_{\mathrm{max}}} (\rho ) \ll 1$, due to the relation
\begin{equation}
 \left( 1- \rho {V_{\mathrm{max}}} (\rho ) \right) ^{ -1/ \rho {V_{\mathrm{max}} } (\rho )  } \sim e ,
\end{equation}
(\ref{ZGfluVK2}) is expressed as
\begin{equation}
 Z_G = \frac{1}{\Lambda ^{dN}} \max _{ \rho } \left\{ \left( 1- \rho {V_{\mathrm{max}}} (\rho ) \right) \rho ^{-1}  e^{1 -\beta \rho U_{0} -\frac{\beta p}{\rho}} \right\} ^N , \label{ZGfluVKvdW}
\end{equation}
This equation coincides with (\ref{ZGvdW}) if we regard $V_{\mathrm{max}} (\rho ) $ as similar to the excluded volume per one particle $b$.

According to the above discussions, the van der Waals theory can be interpreted as a mean-field theory which is derived from mean-field like evaluation of potential and modification of partition function by the excluded-volume effect. This evaluation is also interpreted as the approximation which ignores $U_K$ in (\ref{Approx1}). Although expression of the excluded-volume effect of the approximation of this paper differs from that of the van der Waals theory, this difference is not essential. Furthermore, these two expression coincides with each other in the case of low-density fluid. From these facts, we can regard the approximation of this paper as an extension of the van der Waals theory which consider Fourier component of potential corresponding to nonzero wave numbers additively.

\end{document}